# Behavioral Model For Live Detection of Apps Based Attack

Misbah Shafi, *Student Member, IEEE*, Rakesh Kumar Jha, *Senior Member, IEEE*, Sanjeev Jain, *Senior Member, IEEE*

*Abstract*— **Smartphones with the platforms of applications are gaining extensive attention and popularity. The enormous use of different applications has paved the way to numerous security threats. The threats are in the form of attacks such as permission control attacks, phishing attacks, spyware attacks, botnets, malware attacks, privacy leakage attacks. Moreover, other vulnerabilities include invalid authorization of apps, compromise on the confidentiality of data, invalid access control. In this paper, an application-based attack modeling and attack detection is proposed. Due to A novel attack vulnerability is identified based on the app execution on the smartphone. The attack modeling involves an end-user vulnerable application to initiate an attack. The vulnerable application is installed at the background end on the smartphone with hidden visibility from the end-user. Thereby, accessing the confidential information. The detection model involves the proposed technique of an Application-based Behavioral Model Analysis (ABMA) scheme to address the attack model. The model incorporates application-based comparative parameter analysis to perform the process of intrusion detection. The ABMA is estimated by using the parameters of power, battery level, and the data usage. Based on the source internet accessibility, the analysis is performed using three different configurations as, WiFi, mobile data, and the combination of the two. The simulation results verify and demonstrates the effectiveness of the proposed model.**

*Index Terms*—Security, attack modeling, intrusion detection, smartphone applications, Application-based Behavioral Model Analysis (ABMA), energy consumed

## I. INTRODUCTION

### A. Motivation

In recent years smartphone application models have explosively increased from personnel to professional applications including education, online shopping, net banking, and healthcare. The platform of these applications has massively increased the threat of attacks by compromising trustworthiness and security capabilities [1]-[3]. Third party application marketing is one of the major threat, wherein interested application can be installed by the end-user. However, the applications from these platforms can prove menaces with the advent of vulnerable breaches. Various attacks were identified that can prove detrimental and have adverse effects on the overall security of the information concerned to the smartphone. The jamming attack is one of the prime issues against time-critical applications. The attack exposes the in transit confidential information to the intruders [4]. Inaudible voice attack manipulates voice controllable device with unnoticeable characteristics while operating modulation technique using ultrasonic carriers [5]. The camera-based attack proves a serious security threat to the multimedia applications of smartphones [6]. The side-channel attack exploits the leakage data to limit the data confidentiality on smartphones [7],[8]. Pin inference attack is identified as the privacy threat for the devices controlled by smartphones [9]. Indirect eavesdropping attack is another possible menace that makes use of acoustic sensing to execute the attack on the smartphone [10].

Permission control is one of the primary countermeasures against the possible security risks in smartphones. The permission control enhances the security by incorporating conditional restrictions on the particular executions performed by the applications. Various permission control methodologies were formulated including context sensitive permission control [11], user driven access control [12], permission control using crowd sourcing [13], and Sig PID (Significant Permission Identification) [14]. However, the major limitation associated with the permission control technique is that the targeted functionality of the application is restricted such that the desirable and undesirable private data transmission is not well differentiated.

Data privacy control scheme is another security enhancement technique against application attacks. Seivedroid is the privacy control technique to mark objectionable and confidential data [15]. Privacy-preserving data encryption of applications involves selective data encryption with the association of time constraints [16]. Flow intent is defined as the identification of non-functional confidential data transmission and prevents suspicious data transmissions from application visual interfaces [17]. However, locating sensitive data is one of the major challenges evolved in these mechanisms. Speed and adaptability are the other issues that limit the security enhancement of smartphone applications.

The overall constraints of the conventional security enhancement schemes of smartphone applications include particular application specifications-based security enhancement schemes, detection of sensitive data, requirement

Misbah Shafi is with the department of electronics and communication, Shri Mata Vaishno Devi University, Katra, India 182230, Rakesh Kumar Jha is with the Indian Institute of Information Technology Design and Manufacturing Jabalpur, India 482005, and Sanjeev Jain is with the Central University of Jammu, Jammu and Kashmir, India 181143. (email: misbahshafi0@gmailcom, jharakesh.45@gmail.com, dr_sanjeevjain@yahoo.com)

TABLE I
COMPARISON OF RECENT SCHEMES WITH THE PROPOSED SCHEME

| Ref. | Objective | Attack type | Applied application | Technique/tool | Optimized parameters | Motivation |
|---|---|---|---|---|---|---|
| [18] | Earlier application version as attack threat | Man in the middle attack | Facebook, Sina weibo, qihoo 360 | Driodskynet | Success rate, protection percentage, similarity index of early and upgraded versions | **Disadvantages of [18]-[24]:** Application specific, attack, technology dependent, complex, not applicable to latest or upcoming applications, inadaptive |
| [19] | Communication based attack detection using intent abnormality analysis | Communication based attack, hijacking attack | 57 real world applications | Component level data flow analysis | Detection rate, time overhead, space overhead | |
| [20] | Self defending mechanism to let the repackaged application expose themselves automatically | Reuse attacks / repackaging attacks | RandomMusic, standup timer, angle, asquare, swiFTP, pdfview, opensudoku, achartengine, wordpress, Apidemos | Self defending coding technique | Performance overhead, diversity test | |
| [21] | Antiphishing scheme for mobile platforms | Phishing attack | Facebook, paypal, ebay | MobiFish | Execution time | **Advantages of the proposed scheme** Generalized framework, technology independent, simple, applicable to latest or upcoming applications, adaptive, dependency on live data, live detection of attack |
| [22] | End to end caller ID verification scheme | Caller ID spoofing attack | Caller Dec application | CallerDec | Accuracy, time, voltage usage, precision, recall | |
| [23] | Active warden attack is detected using integrity and multiparty verification framework | Active warden attack | Instagram, frozen bubble | Multiparty verification | Average overhead, code bloat | |
| [24] | Repackage proofing framework for detection and response site | Privacy leakage attack | Frozen bubble | Android data flow analyzer | Average overhead, code bloat | |
| Proposed | Mobile application-based attack model and detection model for mobile application | Mobile application-based attacks | Facebook, whatsapp, gmail, amazon, youtube, chrome | Application thermal pattern analysis | Energy efficiency, power usage, battery usage, data | |

of desirable private data transmission, and independent of updates. To make the security enhancement techniques devoid of these limitations, an application-based attacking model followed by the detection of intruder applications has been proposed.

*B. Related Work*

The current security improvement techniques in view of smartphone application platforms are briefly summarized in Table I. In [18], earlier version applications have been discovered as the source of vulnerable threats of an attack. To counteract the attack possibility, Driodskynet has been developed as a tool to find out and evaluate the applications with security risks from the application installation source such as playstore. In [19], the possible security menaces are located in the android operating system having inter-component communication. The component-level data flow analysis technique has been executed to recognize the caller and the callee on the basis of the data dependencies. However, the communication based attacks are identified by the parameter of the intent abnormality. In [20], a self-defending mechanism has been formulated to allow the repackaged applications to manifest automatically. The scheme encrypts the portion of the application code during the compile-time and the ciphertext code is decrypted at the run time. In [21], an antiphishing scheme MobiFish has been proposed for smartphone platforms. The strategy involves the validity verification of applications, webpages, and other persistent accounts. The validation is obtained by comparing the claimed identity with the actual identity. In [22], end to end caller ID verification technique has been devised by evaluating the current smartphone network infrastructure. A CallerDec application has been designed as an ID spoofing detection tool for android based smartphones to evaluate validation and effectiveness of the mechanism.

*C. Novelty*

Application-based Behavioral Model Analysis (ABMA) is a novel methodology defined for security improvement of smartphone platforms. Conventional schemes of security enhancement mechanisms in smartphones are attack-specific or application-specific. A generalized security scheme independent of versions and type of application is yet to be addressed. Also, the reliability with upgradation and optimized statistical parameters require immense attention. The behavioral model for smartphone based applications is an innovative initiative to address these challenges with an effectual performance. The model is independent of the technology and the updates of the applications of the smartphone. It provides the live detection app based attack with adaptive capability. The detection model ensures the apps based attack detection on authorization, confidentiality, and integrity with efficient and less complex methodology. The significant contribution of this paper is as follows

- A novel and probable application-based attacking model has been identified with an efficient strategy of hidden access installation in the background and the hidden visibility from the end-user.
- The detection model for the applications of the smartphone has been proposed to address the modeled attack. The evaluation is based on the comparative analysis of the behavioral model such that the actual parameters are compared with the parameters in presence of the intruder application.
- To counteract the detected intrusion, an alarm is raised as

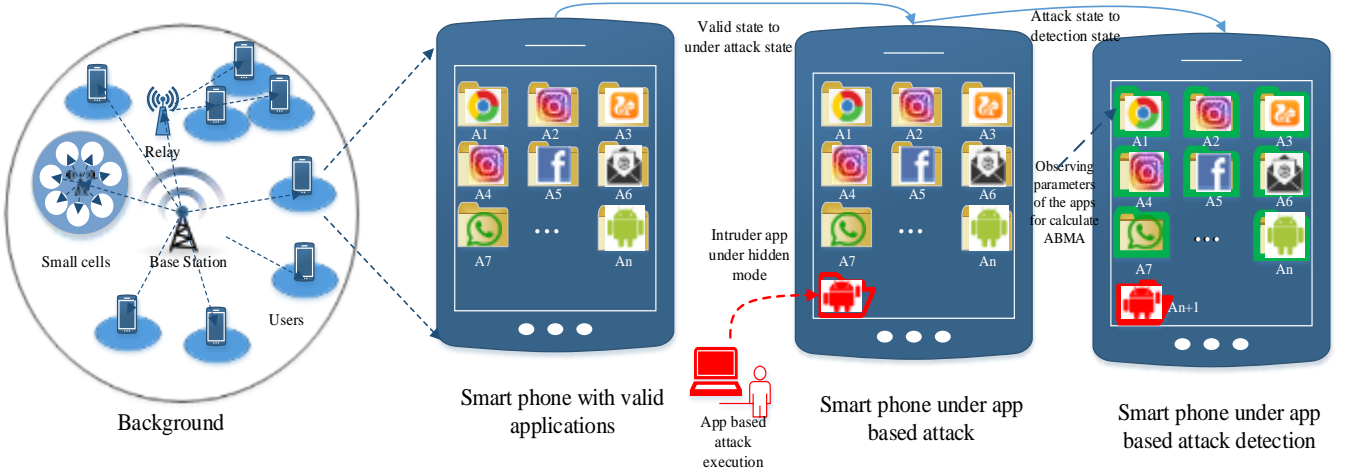

Fig. 1. System model of application-based attack modeling and detection

the immediate response followed by the disconnection of the cellular services and internet accessibility.

- The obtained results illustrate that the proposed scheme can prove an effective mechanism using ABMA in terms of power, data, and battery level.

*D. Organization*

The paper is organized as follows. Section I presents the introduction. Section II describes the system model and problem formulation. The realization and representation of the proposed methodology is discussed in Section III. An illustrative example of the proposed methodology is demonstrated in Section IV. The performance evaluation is analyzed in Section V. To the closure of the paper, the conclusion is given in Section VI.

## II. SYSTEM MODEL AND PROBLEM FORMULATION

In this section, the system model, characteristics model is illustrated for the detailed network description. Further, the problem formulation is discussed for the proposed scheme

*A. System model*

The fundamental of the system model consists of the end-user service application-based scenario. The overall background communication network involves a BS and $i$ number of users. The end-user connectivity services are assumed to be provided by any of the 5G/6G technologies. The $i^{th}$ user is considered as the target of attack, therefore analysed for attack modeling and attack detection. The end-user is portrayed in the form of a smartphone having $n$ number of different applications, where $n \in \mathbb{N}$. Consider $A = A_1, A_2, A_3, \ldots, A_n$ number of availed applications on the smartphone. The intruder is considered to be present in the scenario having an interest of obtaining the confidential information of the valid user from the smartphone. The intruder applies advertisement access add on the other applications to breach through the smartphone validations. The intruder is assumed to have a keen watch on the actions of the valid user on the smartphone. Under any immediate action of the advertisement add the link of advertisement follows to the downloading application such as play store for the download and installation process such that

$$A = \{A_1, A_2, A_3, \ldots, A_n, A_{n+1}\} \quad (1)$$

The $A_{n+1}$ application processes in the hidden visibility mode to prevent the detection. The applications $A = A_1, A_2, A_3, \ldots, A_n$ are present in the list of smart phone. The downloaded applications are more prone to vulnerable breaches and have more compromised security as compared to the system applications. Therefore, we consider the downloaded application as the source link of the vulnerability, advertised on it.

*Proposition 1:* The source background communication connectivity scenario is considered as free from intruders and is completely validated. Thus provides secure internet access with good network coverage from the backend.

*B. Characteristic parameter model*

The parameter of power is one of the parameters that determine the performance of the device. The execution of the attack affects the overall performance thereby increases the power usage. The overall consumed power $P_c$ of the smartphone involves the combination of the static $P_s$ and dynamic power $P_d$ for the time instant $t = 0^+$ to $t = T$ is given by

$$P_c(t) = P_d(t) + P_s(t) \quad (2)$$

Adding the impact of corresponding leakage power. The equation (2) can be re-written as

$$P_c(t) = P_d(t) + P_s(t) + P_{dl}(t) + P_{sl}(t) \quad (3)$$

Where, $P_{dl}$ is the leakage power in the dynamic state, $P_{sl}$ is the leakage power in the static state.

**Definition 1**: The total power consumed in applications $P_{dA}$ is the sum of the power consumed by the individual applications for the time instant $t = 0^+$ to $t = T$. The attack via an application affects the total power consumed by the applications.

Using equation (1), the total power consumed for the corresponding applications can be defined as

$$P_{dA}(t) = P_{A1}(t) + P_{A2}(t) + P_{A3}(t) + \cdots + P_{An}(t) + P_{ckt}(t) \quad (4)$$

*Proposition 2:* For the case of simplicity the circuit power $P_{ckt}$ is excluded for the analysis such that, solely power consumed for the communication is considered.

Based on the demand of the applications and the resource allocation of the corresponding application, the total estimated power for the applications can be obtained as

$$P'_{dA}(t) = \sum_{i=1}^{n} P_{Ai}(t) \quad (5)$$

Or

$$P'_{dA}(t) = \sum_{i=1}^{n} \frac{\varpi_i^2(t)}{|h_{Ai}(t)|^2} \left\{ 2^{\frac{\gamma_i(t)}{\beta_i(t)}} - 1 \right\} \quad (6)$$

Where $\varpi$ is the noise power, $h$ is the channel coefficient, $\beta$ is the allocated bandwidth, $\gamma$ is the data rate of the corresponding application. $P_{Ai}$ is the power consumed by the $ith$ application.

**Definition 2**: The total data rate is the sum of data rates of individual applications. The presence of an attack has a direct impact on the usage of the data.

Using equation (1), the corresponding total data rate $\gamma_A$ of the applications for the time instant $t$ is given by

$$\gamma_A(t) = \sum_{i=1}^{n} \gamma_{Ai}(t) \quad (7)$$

Or
$$\gamma_A(t) = \sum_{i=1}^{n} \beta_i(t) \log_2(1 + \varphi_i(t)) \quad (8)$$

Where $\beta_i \in \{\beta_{A1}, \beta_{A2}, \beta_{A3}, \dots, \beta_{An}\}$ is the corresponding bandwidth allocated to the application, $\varphi$ is the SNR.

**Definition 3**: The total battery consumed in applications is the sum of battery consumed by the individual applications

$$B_A(t = T) = \sum_{i=1}^{n} [B_{Ai}(t = 0^+) - B_{Ai}(t = T)] \quad (9)$$

Where $B_A(t = T)$ defines the total consumed battery level by the applications, $B_{Ai}(t = 0^+)$ is the initial battery level for the $ith$ application, $B_{Ai}(t = T)$ is the final battery level at the time instant $T$ for the $ith$ application. The battery lifetime [23] can be evaluated as

$$L_b(t) = \frac{B_c(t)}{\sum_{i=1}^{n} I_{Ai}(t)} * 0.7 \; (in \; hours) \quad (10)$$

Where $L_b$ is the total battery lifetime, $B_c$ is the total battery capacity available, $I_{Ai}$ is the total current drawn by the $ith$ application. The value 0.7 accounts for external factors.

**Definition 4**: The total energy efficiency provided by the smartphone for the applications is the sum of the individual energy efficiencies of the corresponding applications. Similarly, presence of the attack affects the energy efficiency. For each of the applications, the total energy efficiency of the smartphone for the applications is given by

$$EE_A(t) = \sum_{i=1}^{n} EE_{Ai}(t) \quad (11)$$

Equation (11) can be obtained as

$$EE_A(t) = \sum_{i=1}^{n} \left( \frac{\gamma_{Ai}(t)}{P_{Ai}(t)} \right) \quad (12)$$

These parameters critically define the characteristic behavior of the applications and eventually effects the performance of the smartphone.

**Pseudo Code 1: Behavioral model analysis**

1: **Input parameters**
Battery usage
Battery status
Data rate
Power consumption status
Energy consumption status
Number of applications
Operating bandwidth $\beta$, Noise power $\varpi$

2: **Initialization**
Initialize time $t$
Set of applications $A = \{A_1, A_2, A_3, \dots, A_n\}, n \in \mathbb{N}$
Circuit power consumption $P_{ckt}$
Initial battery level in terms of battery lifetime $L_b$
Initialize power consumption of applications
Initialize energy consumption of applications
Initialize data rate of applications
Initialize battery usage of applications

3: **Intruder application detection using behavioral model analysis**
*Estimated analysis for time instant t*
**for** $i = 1: n$ /*$n$ is the number of applications*/
Determine power consumed by the $ith$ application $P_{dA}(i) = \{P_{Ai}\}$
Determine data rate of the $ith$ application $\gamma_A(i) = \{\gamma_{Ai}\}$
Determine battery usage of the $ith$ application $B_A(i) = \{B_{Ai}\}$
**end**
Compute total power consumed by $n$ applications
$$P'_{dA}(t) = \sum_{i=1}^{n} P_{Ai}(t)$$
Compute total data rate consumed by $n$ applications
$$\gamma_A(t) = \sum_{i=1}^{n} \gamma_{Ai}(t)$$
Compute total battery consumed by $n$ applications
$$B_A(t = T) = \sum_{i=1}^{n} [B_{Ai}(t = 0^+) - B_{Ai}(t = T)]$$
*Observed analysis for time instant t*
**find**
Total power consumed, data rate consumed, battery consumed by the applications $P_{dA}^{obs}(t), \gamma_A^{obs}(t), B_A^{obs}(t)$
**if** $P'_{dA}(t) \neq P_{dA}^{obs}(t), \gamma_A(t) \neq \gamma_A^{obs}(t), B_A(t) \neq B_A^{obs}(t)$
Intruder application is present
Raise an alarm
Disconnect the cellular and internet accessibility
**else**
Secure mobile computing is processing
**end if**
**wait for time instant $\delta t$** /* $\delta t$ is the very small time instant such that constant monitoring is maintained*/

4: **Updation in the number of applications**
**if** $n(t) \neq n(t + \delta t)$
update $n$
**end if**

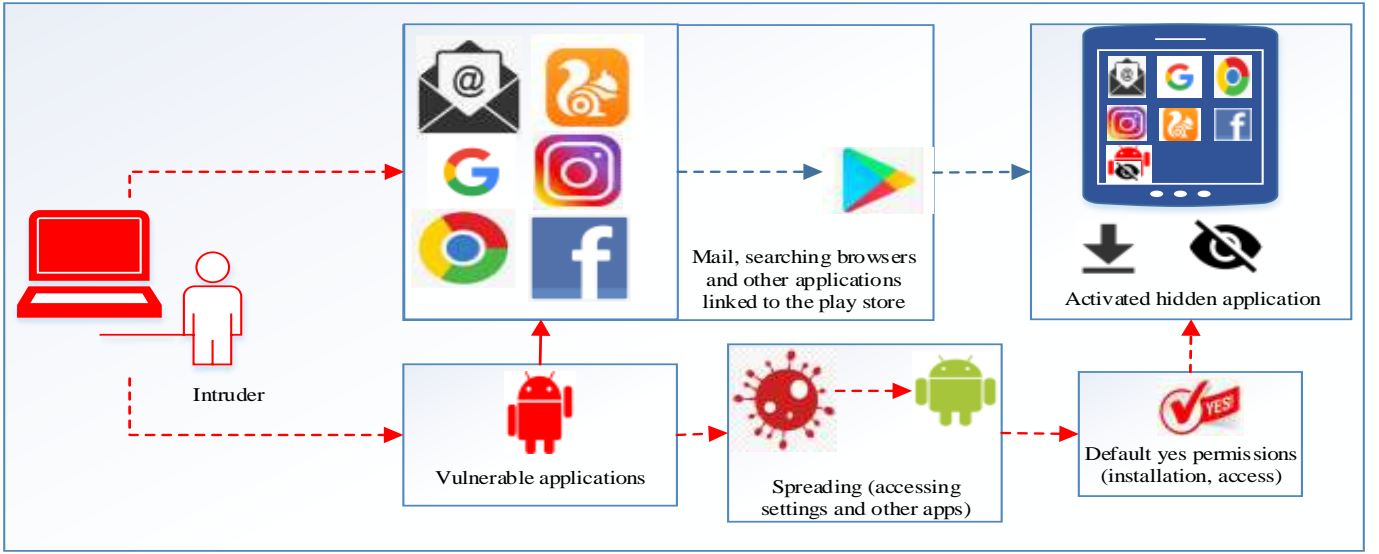

Fig. 2. Apps based attack model

## C. Problem formulation

The proposed scheme identifies the possible attack modeling and intrusion detection based on the characteristic parameters of data rate, energy efficiency, power, and battery. The prime aim is to detect the application-based intrusion using the analysis of these parameters. The collective performance of these parameters define the criteria as

$$P_{es}(t) \approx P_{obs}(t) \qquad (13)$$

Such that,

$$P_{obs}(t) \approx (P_{A1}(t) + P_{A2}(t) + P_{A3}(t) + \cdots + P_{An}(t)) \qquad (14)$$

The corresponding criteria of data rate for intrusion detection is defined as

$$\gamma_{es}(t) \approx \gamma_{obs}(t) \qquad (15)$$

Similarly,

$$B_{es}(t) \approx B_{obs}(t) \qquad (16)$$

Also,

$$E_{es}(t) \approx E_{obs}(t) \qquad (17)$$

Where $P_{obs}, \gamma_{obs}, B_{obs}, E_{obs}$ are the sum of the individual application parameters observed. $P_{es}, \gamma_{es}, B_{es}, E_{es}$ are the total estimated application parameters. Constraints are given in equations (13), (15), (16), and (17) to confirm the validation of the applications while allocating the services. It defines the collective performance of the applications such that the estimated boundary bound value of the parameters is not exceeded.

## III. REALIZATION AND REPRESENTATION

This section presents the realization and representation of the proposed mechanism by depicting a pseudo code-1. The pseudo code-1 provides complete insight of the execution and procedure of intruder application detection. The $n$ valid applications such that $n \in \mathbb{N}$, for time instant $t$ are considered for the behavioral model analysis depicting power, battery, and data rate parameters. The parameters are analyzed based on the performance of individual applications. The estimated parameters in absence of the attack act as a threshold for the detection of the intruder application. Similarly, the behavioral model analysis using these parameters in presence of the intruder application is evaluated, forming the observed analysis. The estimated behavioral model in absence of the attack is compared with the observed behavioral model in presence of the attack to detect the intruder application. Further, for secure mobile computing in absence of the attack, the estimated behavioral model is equivalent to the observed behavioral model. The attack model and the detection model is well shown in the Fig. 2. and Fig. 3.

## IV. PROPOSED METHODOLOGY OF APPLICATION-BASED ATTACK DETECTION: HOW IT WORKS

This section defines the proposed methodology of the application-based attack model and the detection model.

### A. Attack model

The attack modeling defines the execution of the attack by making use of the possible application-based vulnerability. The intruder makes use of the vulnerable application to initialize the attack execution. The intruder provides the installation link on the other installed applications of the cellular smartphone in the form of an advertisement or the pop up option similar to apps based phishing attacks. For any response of the user, the vulnerable application starts to download in the background. The already installed application is assumed to be linked to the play store. In other words, during the internet accessibility and the processing of the already installed applications, the vulnerable application breaches via other downloaded installed application targets the access in the database of the cell phone. The vulnerable application has the ability to perform the function of spreading. Spreading is defined as, accessing settings and other applications. The vulnerable application is incorporated by the default 'Yes' accessibility to make the procedure devoid of the permissions such as permission control vulnerabilities. The whole process results in the installation of the intruder application under hidden visibility. The hidden visibility is defined as the disability mode where the intruder application is not visible on the user account and the list of applications on the smartphone. The mechanism of attack modeling is shown in Fig. 2.

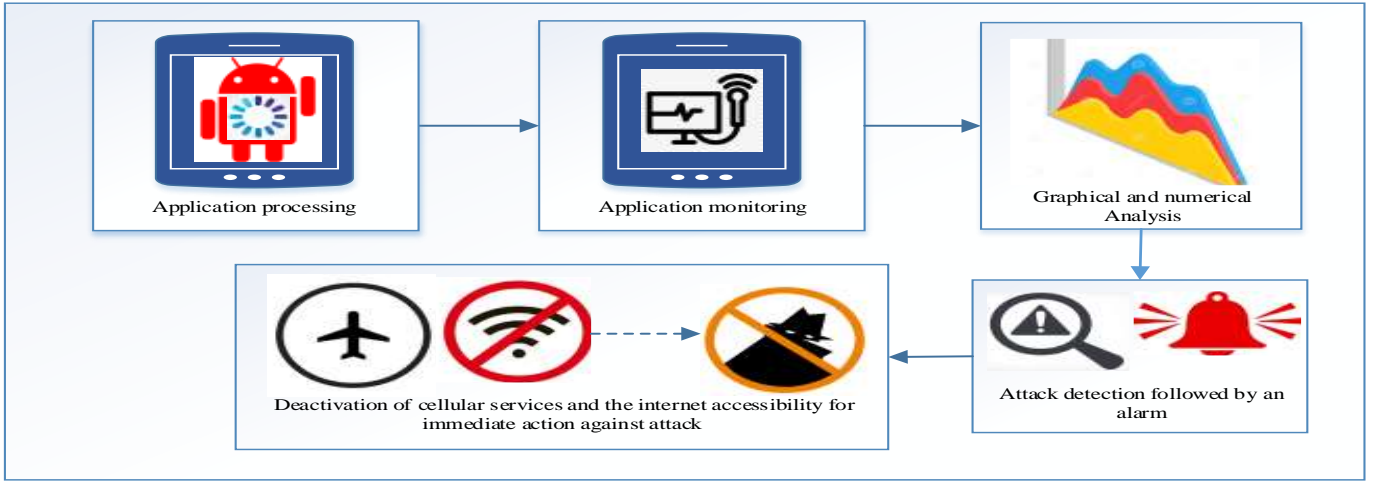

Fig. 3. Apps based detection model

*Illustration:* In absence of the attack, the set of applications is defined as

$$A = \{A_1, A_2, A_3, \ldots, A_n\} \quad (18)$$

The corresponding parameters for different applications are given in equation (5), (7), (10), and (11). In presence of the attack, the set of applications is defined in the equation (1).

*B. Application intrusion detection model*

The detection model carries out the schematic of intrusion detection for application-based attack. During the invalid application processing in the memory and database of the cellular smartphone, the monitoring of each application is observed using ABMA. The mechanism of ABMA is performed based on the power consumption, battery level, and data usage. Other parameters such as energy efficiency, complexity, battery consumption are parameters that can be considered for the analysis. The ABMA in absence of the intruder acts as a threshold for intruder app detection. The threshold is adaptively obtained from the live data of applications. From the comparative analysis of the ABMA pattern, the normal and the invalid installed application is detected. The attack detection is followed by the alarm or the pop up notification of the invalid app activity. For immediate protection against the attack, the cellular services and the internet accessibility (WiFi, mobile data, BlueTooth) are turned off. The procedure of deactivation of services is informed to the user with the popup notification. The process flow of the intruder app detection mechanism is shown in Fig. 3

*Illustration:* The parameters in absence of attack are considered as the threshold for intrusion detection. The behavioral model characteristics for the intrusion detection can be defined as

$$P_A^{att}(t) > P_A^{val}(t) \quad (19)$$

Similarly,

TABLE II
SIMULATION PARAMETERS

| Parameter | Value |
|---|---|
| Number of valid applications for analysis | 6 |
| Maximum battery capacity | 4000mAh |
| WiFi link speed | 72Mbps |
| Number of intruder application | 1 |
| Mobile Type | android |
| Smartphone specifications | Redmi Xiaomi, Note 7 Pro |
| Mobile brightness | 50% |
| Reading mode | off |
| Android version | 10 QKQ1.190915.002 |
| Battery status | Not charging |
| Average mobile data speed | 1 Mbps |

$$\gamma_A^{att}(t) > \gamma_A^{val}(t) \quad (20)$$

Also,

$$B_A^{att}(t) < B_A^{val}(t) \quad (21)$$

And,

$$E_A^{att}(t) > E_A^{val}(t) \quad (22)$$

Where, $P_A^{att}$ is the total power consumed by the applications in presence of the attack. $P_A^{val}$ is the total power consumed by the applications in absence of the attack. $\gamma_A^{att}$ is the total data rate of the applications in presence of the attack. $\gamma_A^{val}$ is total data rate of the applications in absence of the attack. $B_A^{att}$ is the battery level in presence of the attack, $B_A^{val}$ is the battery level in absence of the attack. $E_A^{att}$ is the total energy consumption by the applications in presence of the attack, $E_A^{val}$ is the total energy consumption by the applications in absence of the attack. (Refer Lemma 1) ∎

V. SIMULATION ANALYSIS

This section provides the complete insight of the analytical results attained by the real-time investigation of the apps based attack. The scenario involves the apps based analysis on the smartphone. The specifications of the analysis parameters are shown in Table II. The section is divided into two parts. The first defines the ABMA and the second part evaluates the results

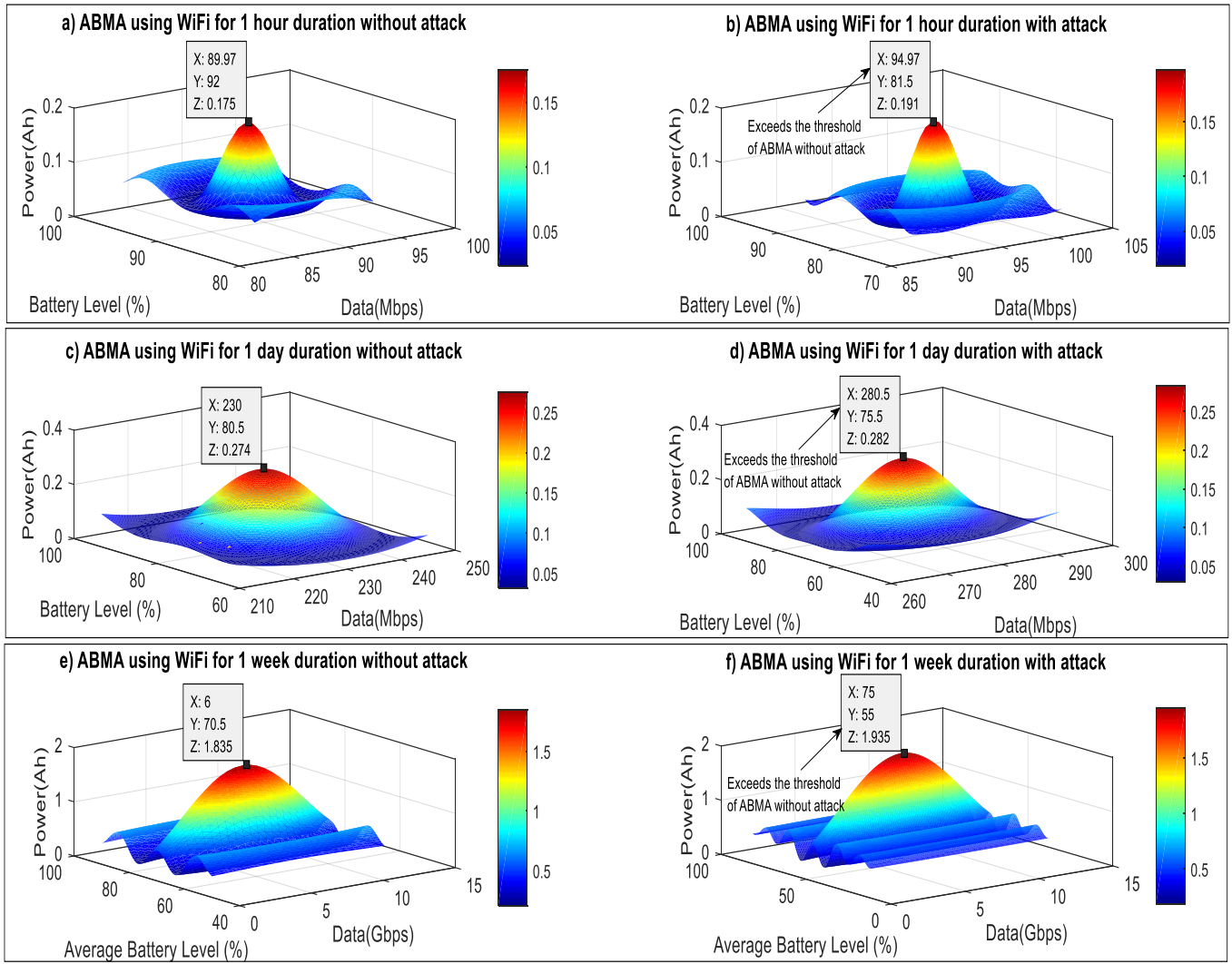

Fig. 4. Real-time Application-based Behavioral Model Analysis (ABMA) for attack detection

TABLE III
REAL TIME ANALYSIS FOR DIFFERENT APPLICATIONS USING MOBILE DATA

| Duration | Application type | Average Data (Mbps) | Average Battery usage (%) | Average Power consumed (mAh) | Valid /invalid app | Duration | Application type | Average Data per day (Mbps) | Average Battery usage /day(%) | Average Power consumed (mAh) | Valid /invalid app |
|---|---|---|---|---|---|---|---|---|---|---|---|
| 1 hour | Whatsapp | 0.577 | 2 | 16.44 | Valid | 1 week | Whatsapp | 12.06 | 5 | 208.62 | Valid |
|  | Facebook | 32.6 | 7.23 | 127.150 | Valid |  | Facebook | 100.1 | 8.2 | 307.19 | Valid |
|  | Youtube | 42.9 | 9.1 | 160.037 | Valid |  | Youtube | 299.5 | 30.19 | 848.61 | Valid |
|  | Chrome | 12.5 | 6.74 | 118.53 | Valid |  | Chrome | 128.26 | 10.2 | 283.08 | Valid |
|  | Gmail | 0.241 | 0.83 | 14.596 | Valid |  | Gmail | 9.4 | 3.87 | 129.21 | Valid |
|  | Amazon | 0.545 | 1.98 | 34.82 | Valid |  | Amazon | 10.61 | 4 | 164.82 | Valid |
|  | Intruder application | 30.7 | 5.22 | 174.15 | Invalid |  | Intruder application | 96.6 | 28.2 | 598.16 | Invalid |
| 1 day | Whatsapp | 1.3 | 4.41 | 72.81 | Valid | 1 month | Whatsapp | 66.2 | 12.24 | 510.72 | Valid |
|  | Facebook | 57.5 | 20.17 | 354.72 | Valid |  | Facebook | 157.7 | 10 | 374.626 | Valid |
|  | Youtube | 86.6 | 25 | 443.4 | Valid |  | Youtube | 322.13 | 22 | 974.03 | Valid |
|  | Chrome | 76.7 | 25.6 | 454.041 | Valid |  | Chrome | 302 | 20.11 | 558.12 | Valid |
|  | Gmail | 4.7 | 5 | 85.3 | Valid |  | Gmail | 58.7 | 8 | 344.58 | Valid |
|  | Amazon | 1.1 | 3 | 49.53 | Valid |  | Amazon | 101.7 | 3.1 | 127.74 | Valid |
|  | Intruder application | 52.2 | 20.23 | 362.15 | Invalid |  | Intruder application | 112 | 7 | 230.63 | Invalid |

TABLE IV
REAL TIME ANALYSIS FOR DIFFERENT APPLICATIONS USING HYBRID (MOBILE DATA+WIFI) INTERNET SOURCE ACCESSIBILITY

| Duration | Application type | Average Data (Mbps) | Average Battery usage (%) | Average Power consumed (mAh) | Valid /invalid app | Duration | Application type | Average Data per day (Mbps) | Average Battery usage /day(%) | Average Power consumed (mAh) | Valid /invalid app |
|---|---|---|---|---|---|---|---|---|---|---|---|
| 1 hour | Whatsapp | 0.5016 | 1.80 | 14.8 | Valid | 1 week | Whatsapp | 13.23 | 5.14 | 171.44 | Valid |
| | Facebook | 56.2 | 8 | 65.17 | Valid | | Facebook | 112.3 | 11.32 | 232.48 | Valid |
| | Youtube | 17.7 | 7 | 53.05 | Valid | | Youtube | 192.68 | 12.35 | 162.05 | Valid |
| | Chrome | 2.7 | 3 | 64.19 | Valid | | Chrome | 54.92 | 6.026 | 165.76 | Valid |
| | Gmail | 0.372 | 0.8 | 14.8 | Valid | | Gmail | 11.24 | 3.556 | 114.48 | Valid |
| | Amazon | 1 | 1.41 | 26.16 | Valid | | Amazon | 24.4 | 2.27 | 51.883 | Valid |
| | Intruder application | 21.1 | 4.73 | 166.81 | Invalid | | Intruder application | 109 | 11..61 | 259.6 | Invalid |
| 1 day | Whatsapp | 1.6 | 3.33 | 61.59 | Valid | 1 month | Whatsapp | 78.6 | 10.29 | 437.95 | Valid |
| | Facebook | 72.3 | 17.6 | 312.16 | Valid | | Facebook | 202.2 | 8.36 | 320.12 | Valid |
| | Youtube | 44.8 | 13.07 | 241.87 | Valid | | Youtube | 358.8 | 17 | 752.66 | Valid |
| | Chrome | 8.8 | 5.82 | 107.66 | Valid | | Chrome | 241.2 | 9.26 | 325.44 | Valid |
| | Gmail | 1.3 | 3.57 | 70.2 | Valid | | Gmail | 66.9 | 6.3 | 258.44 | Valid |
| | Amazon | 3.3 | 4.10 | 75.92 | Valid | | Amazon | 128.5 | 1.3 | 53.57 | Valid |
| | Intruder application | 61.5 | 23.7 | 253.6 | Invalid | | Intruder application | 215 | 10.25 | 328.8 | Invalid |

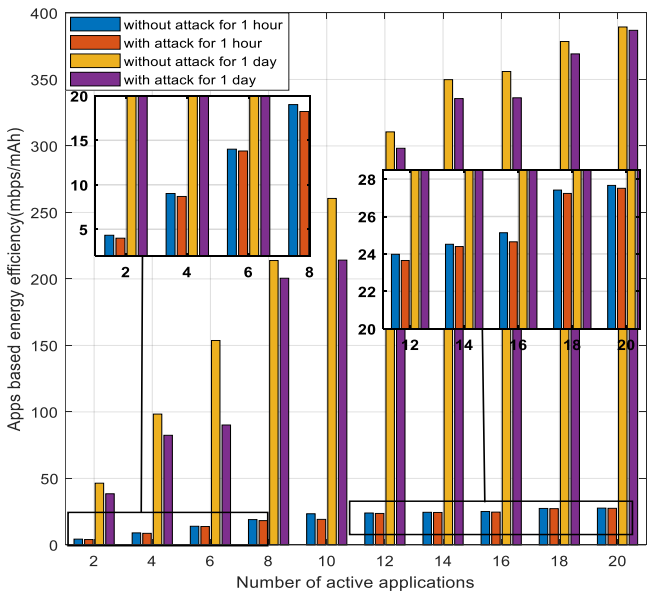

Fig. 5: Apps based energy efficiency analysis with respect to the number of active applications

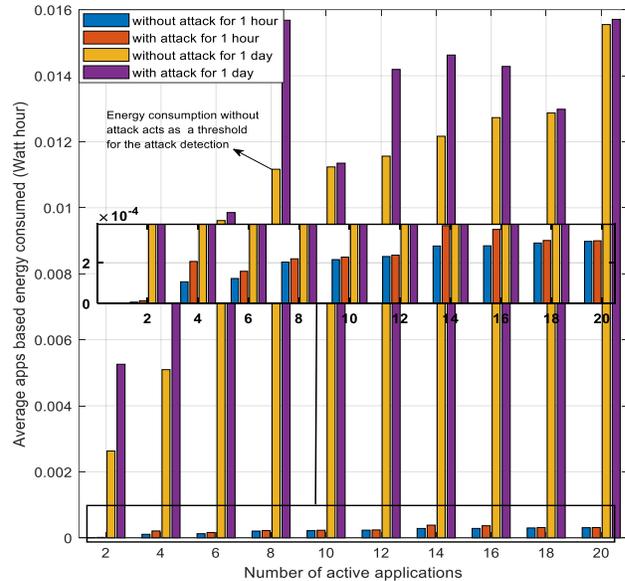

Fig. 6: Apps based energy consumed with respect to the number of active applications

### A. Apps based intrusion detection using ABMA

The radiation pattern demonstrates the comparative Application-based Behavioral Model Analysis (ABMA) for the apps based intrusion detection. The behavioral model determines the apps based peak power usage corresponding to the data and battery level. Fig. 4(a) demonstrates the 3D ABMA pattern for the 1-hour duration using WiFi access in absence of the intruder. A comparable rise in the power, data, and the decrease in battery level is observed in presence of the intruder. The power, data, and the battery level incorporated by the intruder app for its processing ultimately affect the overall usage as shown in Fig. 4(b). Therefore, intruder app processing consumption forms the basis of the proposed methodology of intrusion detection.

Further, for the one day and one week duration, the strategy of comparative ABMA is followed for intrusion detection as shown in Fig. 4(c), 4(d), 4(e), and 4(f).

Table III provides the complete understanding of the ABMA using six different applications using mobile data as the source of internet access for the apps on the smartphone. The comparative corresponding intruder app processing consumption forms the criterion of the intrusion detection. A minor processing activity by the intruder app leads to the change in the final value of the ABMA.

Table IV depicts the ABMA parameter analysis using the combination of both the mobile data and the WiFi for the internet access of the app processing.

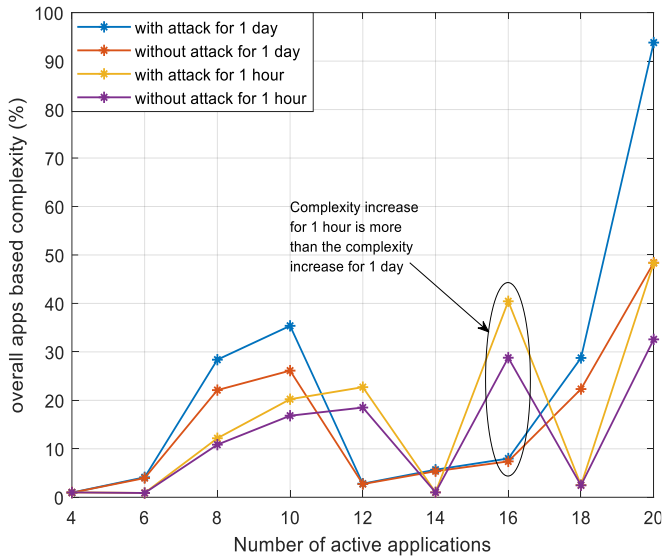

Fig. 7: Complexity with respect to the number of active applications

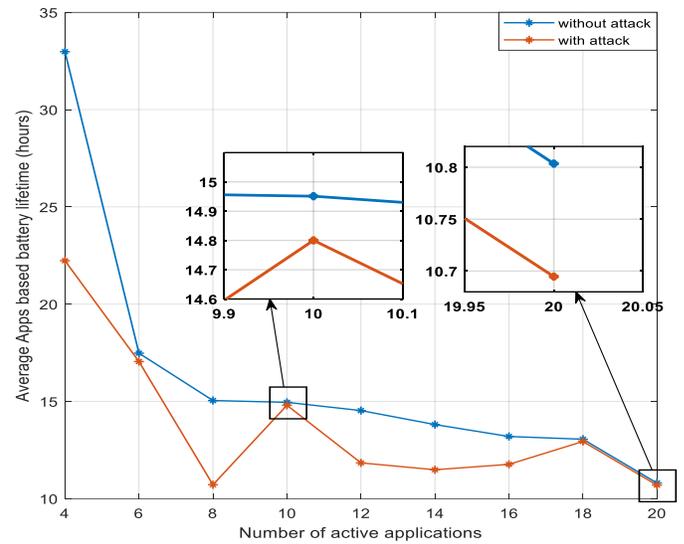

Fig. 9: Battery lifetime with respect to the number of active applications

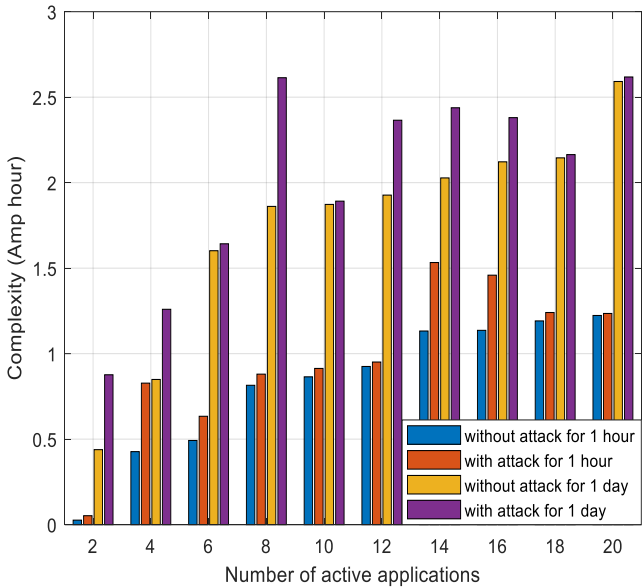

Fig. 8: Complexity with respect to the number of active applications

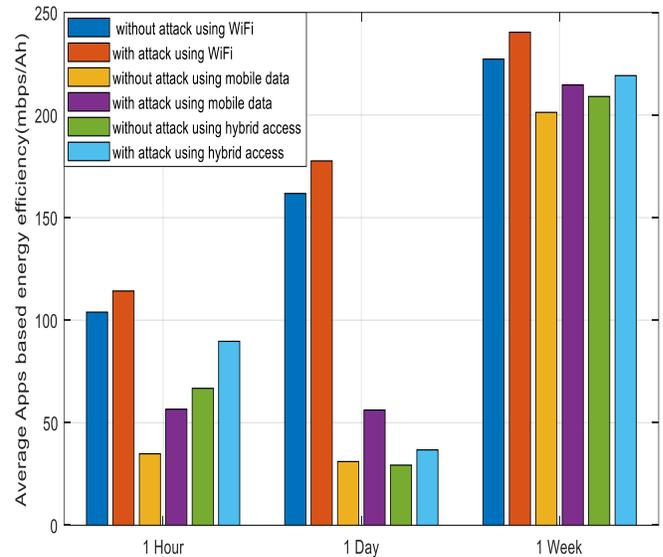

Fig. 10: Apps based energy efficiency analysis

### B. Results and implications

Fig. 5. illustrates the apps based energy efficiency analysis with respect to the number of active applications. It has been observed that with an increase in the number of active applications, the apps based energy efficiency tends to increase thereby, showing a direct relationship with active applications of the smartphone. Thus, more applications are executed with the same amount of energy. In presence of the intruder, the apps based energy efficiency reduces contrary to the valid application-based energy efficiency. The presence of the intruder increases the comparable energy consumption thereby deteriorating the energy efficiency. However, the impact of the intruder is dominant for the one day duration as compared to the one hour duration.

Fig. 6. represents the effect of the intruder app on the parameter of energy consumption. From the analysis, it is observed that with an increase in the number of active applications, energy consumption tends to increase. The presence of the intruder has a drastic impact on overall energy consumption. The intruder processing increases the requirement of energy. Consequently, creates a rise in overall energy consumption. Thus, the increase in the app-based energy consumption in the presence of the intruder can be used as the criterion for app-based intrusion detection. It is clearly detected that the influence of intruder applications is majorly at the longer durations.

The presence of the intruder increases the utilization of the available resources in the form of energy, CPU usage, current drawn, data, and power. In the presence of more number of active applications more is the processing requirement. Therefore, more is the complexity. Moreover, resource utilization has a direct impact on the duration. Resource utilization increases with an increase in the duration of the active applications. Thus, increases the overall app-based complexity. The complexity in terms of the percentage increase is depicted in Fig. 7. The complexity in terms of power usage is shown in Fig. 8.

Fig. 9. provides a clear understanding of the battery lifetime with respect to the increase in the number of active applications of the smartphone. It is illustrated in the figure that the battery

lifetime worsens in the presence of the intruder app. The activation of the intruder app processing upsurges the battery consumption ultimately affecting the overall battery lifetime.

Fig. 10. shows the average app-based energy efficiency under six active apps. Different sources of internet access are compared to demonstrate the parameter of overall energy efficiency with and without the influence of the intruder. The WiFi-based internet source for the apps processing is perceived having the maximum apps-based overall energy efficiency. Based on the duration the overall one-week duration provides the maximum apps based energy efficiency. Hybrid access designates the alternate combination of WiFi internet access and mobile data-based internet access. Hybrid access provides higher apps based energy efficiency as compared to mobile data access. A radical difference is observed between the presence and absence of the intruder app.

*C. Limitations*

The proposed model can be extended using artificial intelligence associated with the mechanisms of reward and punishment such as reinforcement learning algorithms. Artificial intelligence can prove effectual in terms of accuracy, precision, and reliability. However, the introduction of artificial intelligence in the proposed model increases the requirement of high-end processors and more complexity. The proposed behavioral model can prove effective in specific applications such as net banking applications, online shopping applications, social networking applications, stock market applications and other applications that require preeminent confidentiality.

## VI. CONCLUSION

The increase in the use of applications on the smartphone has enhanced numerous vulnerabilities and threats in the form of loss in confidentiality, invalid access control permissions, and invalid authorizations, links to vulnerable sources. In this paper, an application-based attack modeling and attack detection is proposed to address such challenges. The attack modeling incorporates the end-user vulnerable application installation on the smartphone. The possible installation integrates hidden visibility activation mode to process the mechanism. The detection process evaluates ABMA scheme for the invalid application entry. The application-based analysis is estimated using power consumption, battery level, and data usage. The comparative analysis is observed for application intrusion detection. For the immediate countermeasure of the attack, an alarm is raised followed by the disconnection of cellular services and internet accessibility.

## APPENDIX

■ *Lemma 1:* The constraints defined in equations (19), (20), (21), and (22) in terms of power, battery, data rate, and energy specify optimal detection of the vulnerable attack application. The criterion for the application-based attack detection is as follows

  *i.* $P_A^{att}(t) > P_A^{val}(t)$
  *ii.* $\gamma_A^{att}(t) > \gamma_A^{val}(t)$
  *iii.* $B_A^{att}(t) < B_A^{val}(t)$
  *iv.* $E_A^{att}(t) > E_A^{val}(t)$

*Proof*

*i.* We note from equation (5), in absence of the intruder application the total power consumed by the valid applications can be estimated as

$$P_A^{val}(t) = \sum_{i=1}^{n} P_{Ai}(t) \qquad (23)$$

Now in presence of the intruder application, draws some amount of power to execute the processing such that,

$$P_A^{att}(t) = \sum_{i=1}^{n} P_{Ai}(t) + P_{A(n+1)}^{att}(t) \qquad (24)$$

Using equation (23) in equation (24), we get

$$P_A^{att}(t) = P_A^{val}(t) + P_{A(n+1)}^{att}(t), \quad P_{A(n+1)}^{att}(t) \neq 0 \quad (25)$$

Now, the power consumed by the intruder application can be obtained as

$$P_{A(n+1)}^{att}(t) = P_A^{att}(t) - P_A^{val}(t) \qquad (26)$$

From equation (25) and (26), it is concluded that,

$$P_A^{att}(t) > P_A^{val}(t) \qquad (27)$$

*ii.* Part *ii.* follows the similar as that of the part *i.* Using equation (7), in absence of the attack, the total data rate of the valid application for time instant $t$ is obtained as

$$\gamma_A^{val}(t) = \sum_{i=1}^{n} \gamma_{Ai}(t) \qquad (28)$$

Equation (28) is also represented as

$$\gamma_A^{val}(t) = \sum_{i=1}^{n} \beta_i(t) \log_2(1 + \varphi_i(t)) \qquad (29)$$

As per the intruder application requirements, the resource allocation takes place such that the total data rate of the applications in presence of the intruder application is estimated as

$$\gamma_A^{att}(t) = \gamma_A^{val}(t) + \beta_{n+1} \log_2(1 + \varphi_{n+1}(t)) \qquad (30)$$

Equation (30) can be deduced as

$$\gamma_A^{att}(t) > \gamma_A^{val}(t) \qquad (31)$$

*iii.* Part *iii.* specifies the battery evaluation to detect the attack. Using equation (9), the total consumed battery for $n$ valid applications is given as

$$B_A(t = T) = \sum_{i=1}^{n}[B_{Ai}(t = 0^+) - B_{Ai}(t = T)] \qquad (32)$$

Let

$$\sum_{i=1}^{n}[B_{Ai}(t = 0^+) - B_{Ai}(t = T)] = |Z(t)|_{t=0^+}^{t=T} \qquad (33)$$

For the intruder application, the consumed battery for time instant $t$ is denoted as

$$B_{A(n+1)}^{att}(t = T) = [B_{A(n+1)}(t = 0^+) - B_{A(n+1)}(t = T)] \quad (34)$$

Let

$$[B_{A(n+1)}(t = 0^+) - B_{A(n+1)}(t = T)] = |Y_{n+1}(t)|_{t=0^+}^{t=T} \quad (35)$$

Noting from equation (10), the battery life time in absence of the attack is given as

$$L_b(t) = \frac{B_c(t)}{\sum_{i=1}^{n} I_{Ai}(t)} * 0.7 \ (in \ hrs) \qquad (36)$$

The remaining battery lifetime (battery level) after the use of $n$ applications for time $T$ can be interpreted as

$$L_b^r(t) = \frac{B_c(t) - |Z(t)|_{t=0^+}^{t=T} - B_{ckt}(t)}{\sum_{i=1}^{n} I_{Ai}(t) + I_{ckt}} * 0.7 \ (in \ hrs) \qquad (37)$$

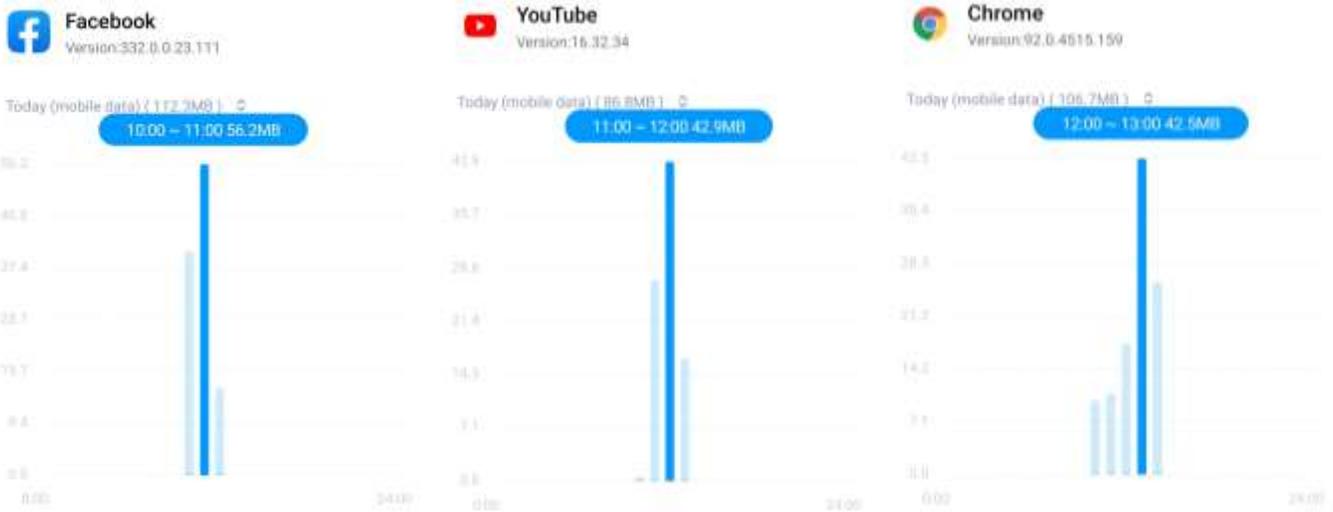

Fig. 11. Real-time application-based analysis using mobile data access for one day analysis.

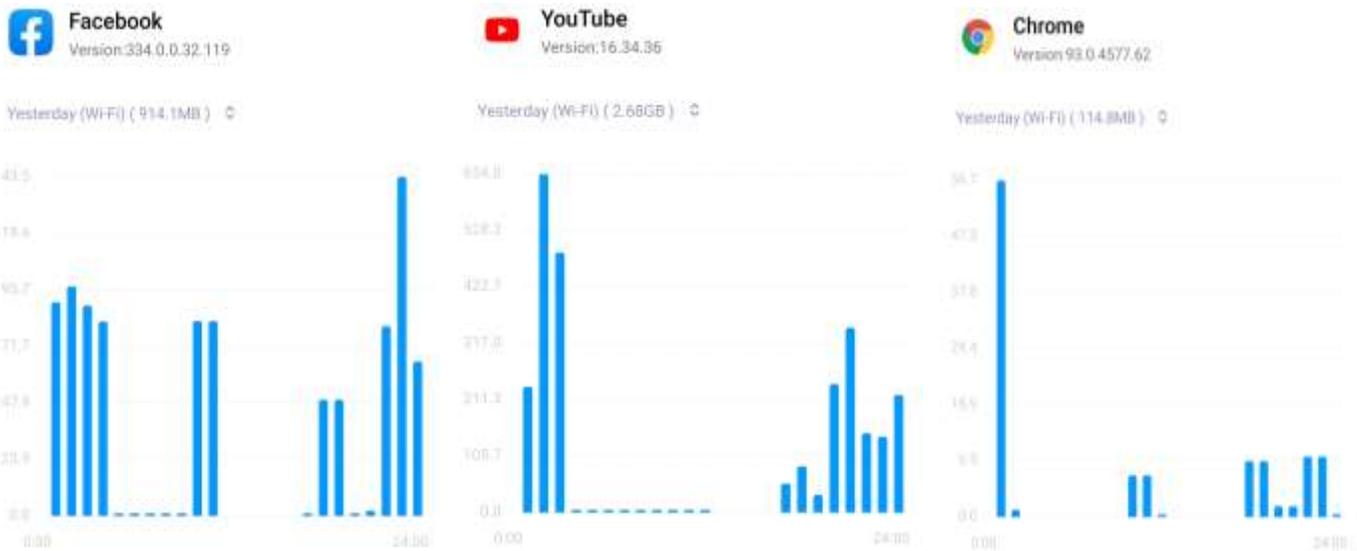

Fig. 12. Real-time application-based analysis using WiFi access for one day analysis.

Where $B_{ckt}(t) = C$ is the battery consumed in processing other than applications and is considered as a constant, $I_{ckt} = D$ is the current drawn in other processing. The equation (37), can be denoted as

$$L_b^r(t) = \frac{B_c(t) - |Z(t)|_{t=0^+}^{t=T} - C}{\sum_{i=1}^{n} I_{Ai}(t) + D} * 0.7 \ (in \ hrs) \quad (38)$$

Now, using equation (35), under the attack in presence of intruder application the battery lifetime can be estimated as

$$L_b^{att}(t) = \frac{B_c(t) - |Z(t)|_{t=0^+}^{t=T} - C - |Y_{n+1}(t)|_{t=0^+}^{t=T}}{\sum_{i=1}^{n} I_{Ai}(t) + D + I_{A(n+1)}(t)} * 0.7 \ (in \ hrs) \quad (39)$$

Comparing equation (37) and (39), we get

$$L_b^r(t) > L_b^{att}(t) \quad (40)$$

Correspondingly, from equation (40), it can be deduced as

$$B_A^{val}(t) > B_A^{att}(t) \quad (41)$$

*iv.* The energy parameter analysis specifies intruder application detection. Noting from equation (12) in absence of the intruder application the overall energy efficiency is given as

$$EE_{Tot}^{val}(t) = \frac{\gamma_{Tot}^{val}(t)}{P_{Tot}^c} \quad (42)$$

Now, in presence of the intruder application, the energy efficiency is obtained as

$$EE_{Tot}^{att}(t) = \frac{\gamma_{Tot}^{val}(t) - \gamma_{A(n+1)}^{att}(t)}{P_{Tot}^c + P_{A(n+1)}^c} \quad (43)$$

Comparing equation (42) and (43), we get

$$EE_{Tot}^{val}(t) > EE_{Tot}^{att}(t) \quad (44)$$

Where $P_{Tot}^c$ is the total power consumed, $\gamma_{Tot}^{val}(t)$ is total available data rate, $\gamma_{A(n+1)}^{att}(t)$ is the allotted data rate to the intruder application, $P_{A(n+1)}^c$ is the power consumed in intruder application.

Similarly, the energy consumed for $n$ applications in absence of the attack can be given as

$$E_T^{val}(t) = \frac{E_A^{Bat}(t) L}{3600} \text{ Watt hr} \quad (45)$$

Where $E_A^{Bat}$ is the battery consumed, $L$ is the transmission latency. In presence of the intruder application the energy consumed is defined as

$$E_T^{att}(t) = \frac{\left(E_A^{Bat}(t) + E_{A(n+1)}^{Bat}(t)\right)L}{3600} \text{ Watt hr} \quad (46)$$

Using equation (45) in equation (46), we get

$$E_T^{att}(t) = E_T^{val}(t) + \frac{\left(E_{A(n+1)}^{Bat}(t)\right)L}{3600} \text{ Watt hr} \quad (47)$$

From equation (47), we get

$$E_T^{att}(t) > E_T^{val}(t) \quad (48)$$

Where $E_{A(n+1)}^{Bat}(t)$ is the battery consumed by the intruder application.

■ Few examples of real-time app analysis is shown in Fig. 11. and Fig. 12.

## Acknowledgment

The authors gratefully acknowledge the support provided by 5G and IoT Lab, DoECE, and TBIC, Shri Mata Vaishno Devi University, Katra, Jammu.

**Misbah Shafi** has received the B.E degree and M.Tech degree in Electronics and Communication Engineering in 2016 and 2018. She is currently pursuing a Ph.D. degree in Electronics and Communication Engineering at SMVD University, J&K. Her research interest includes network security and wireless communication. Currently, she is doing her research on security issues of 5G, 5G NR, and 6G.

**Dr. Rakesh K Jha (S'10, M'13, SM 2015)** is currently an associate professor in the department of Electronics and Communication Engineering, Indian Institute of Information Technology Design and Manufacturing Jabalpur, India. He has also worked as an associate professor at SMVD University, J&K, India. He is among the top 2% researchers of the world. He has published more than 41 Science Citation Index Journals Papers including many IEEE Transactions, IEEE Journal, and more than 25 International Conference papers. His area of interest is Wireless communication, Optical Fiber Communication, Computer Networks, and Security issues. Dr. Jha's one concept related to the router of Wireless Communication was accepted by ITU in 2010. He has received the young scientist author award by ITU in Dec 2010. He has received an APAN fellowship in 2011, 2012, 2017, and 2018 and a student travel grant from COMSNET 2012. He is a senior member of IEEE, GISFI and SIAM, International Association of Engineers (IAENG) and ACCS (Advance Computing and Communication Society). He is also a member of, ACM and CSI, with many patents and more than 1941 citations on his credit.

**Sanjeev Jain** is currently working as a vice chancellor of Central University of Jammu, Jammu and Kashmir, India. He has worked as a director at Indian Institute of Information Technology Design and Manufacturing Jabalpur, India. He has also worked as a vice chancellor at SMVD University, Katra. He has served as Director, Madhav Institute of Technology and Science (MITS), Gwalior. Besides teaching he has the credit of making significant contribution to R&D in the area of Image Processing and Mobile Adhoc Network. His work on Digital Watermarking for Image Authentication is highly valued in the research feld.